\documentstyle[aps,preprint]{revtex}
\begin{document}
\draft
\tighten
\preprint{\vbox{
 \hfill  TIT/HEP-419/NP
 }}
\title{$\pi NN$ coupling determined beyond the chiral limit}
\author{Hungchong Kim \footnote{E-mail : 
hckim@th.phys.titech.ac.jp, JSPS fellow}
}
\address{ Department of Physics, Tokyo Institute of Technology, Tokyo 
152-8551, 
Japan }

\maketitle
\begin{abstract}

Within the conventional QCD sum rules, we calculate the $\pi NN$
coupling constant, $g_{\pi N}$, beyond the chiral limit
using two-point correlation function with a pion.
We consider the Dirac structure, $i\gamma_5$, at $m_\pi^2$ order, which
has clear dependence on the PS and PV coupling schemes for the
pion-nucleon interactions.  For a consistent treatment of the sum rule,
we include the linear terms in quark mass as they
constitute the same chiral order as $m_\pi^2$.
Using the PS coupling scheme for the pion-nucleon interaction,
we obtain $g_{\pi N}=13.3\pm 1.2$, which is very close to the 
empirical $\pi NN$
coupling.
This demonstrates that going beyond the chiral limit is crucial
in determining the coupling and the pseudoscalar coupling
scheme is preferable from the QCD point of view.  
\end{abstract}

QCD sum rule~\cite{SVZ} is a framework which connects hadronic
parameters with QCD parameters.  In this framework, a correlation function
is introduced in terms of interpolating fields constructed from quark and
gluon fields.  The interpolating field is constructed so that its
coupling to the hadron of concern is expected to be strong while its 
couplings to other higher resonances are hoped to be small. Then the 
correlator is
calculated by Wilson's operator product expansion (OPE) in
the deep Euclidean region ($q^2 = -\infty $) and matched with
the phenomenological ``ansatz'' to extract the hadron's information in
terms of QCD parameters.

One interesting quantity to be determined is the pion-nucleon coupling
constant, $g_{\pi N}$.  As the coupling is empirically well-known,
successful reproduction of this quantity may
provide a solid framework to determine other meson-baryon couplings
as well as a better understanding of nonperturbative nature of hadrons.
For this direction, the two-point correlation function for the 
nucleon interpolating field
$J_N$,
\begin{eqnarray} 
\Pi (q, p) = i \int d^4 x e^{i q \cdot x} \langle 0 | T[J_N (x)
{\bar J}_N (0)]| \pi (p) \rangle \ ,
\label{two}
\end{eqnarray}
may be useful and it is often used in calculating 
$g_{\pi N}$~\cite{qsr,hat,krippa1,hung1,hung2}.
Alternative approach is to consider the correlation function without pion
but in an external axial field~\cite{bely1}.  This provides
the nucleon axial charge, $g_A$, which can be converted to $g_{\pi N}$
with the help of the Goldberger-Treiman relation.
Our interest in this work is to provide a reasonable value of 
$g_{\pi N}$ using Eq.~(\ref{two}) because its extension to other meson-baryon
couplings 
seems to be more straightforward.  Further advantage in using
Eq.~(\ref{two}) is to provide a criterion for the PS-PV coupling schemes
for the pion-nucleon interaction as will be discussed below.

The correlation function, Eq.~(\ref{two}),
contains various independent Dirac structures,
each of which can be in principle used to calculate $g_{\pi N}$.
For example, Ref.\cite{krippa1} uses the $\gamma_5 \not\! p$ 
structure while Ref.~\cite{hat} uses the $i\gamma_5$ structure in the
soft-pion limit.
In the recent calculations~\cite{hung1,hung2}, we proposed to use the 
$\gamma_5 \sigma_{\mu \nu}$ structure in studying $g_{\pi N}$ 
as this structure is independent of the pseudoscalar (PS) and 
pseudovector (PV)
coupling schemes employed in the phenomenological side.
This sum rule contains very small contribution from the transition
$N\rightarrow N^*$, and the result is insensitive to the
continuum threshold. 
Therefore, this structure provides a value of $g_{\pi N}$ independent of
the coupling schemes. However, 
the result from this Dirac structure, $g_{\pi N}\sim 10$,
is not quite satisfactory.  
Certainly a further improvement of this sum rule may be needed for the 
future extension to other SU(3) mesons. 

Various improvements can be sought for.  These may include
a question related to the use of Ioffe's nucleon current for the correlator,
higher order corrections in the OPE, or corrections associated
with the chiral limit.  The last possibility for the improvement
is interesting because $g_{\pi N}$ from the $\gamma_5 \sigma_{\mu \nu}$
sum rule
is rather close to the one in the chiral limit than
its empirical value.  In Ref.~\cite{hung1}, the calculation is performed
beyond the soft-pion limit by taking the leading order of the pion
momentum $p_\mu$, but for the rest of the correlator 
the chiral limit, $p^2=m^2_\pi =0$, is 
taken.  Thus, it is not clear whether the calculation is performed
beyond the chiral limit and this may cause the discrepancy
with the empirical $g_{\pi N}$.

In this paper, we pursue an improvement by presenting a QCD sum rule
calculation beyond the chiral limit.  Specifically, we consider the
Dirac structure, $i\gamma_5$, at the order, $p^2=m_\pi^2$.
The sum rule for the structure, $i\gamma_5$, is, first of all,
technically less involved when the calculation is done beyond the chiral
limit.  Secondly, even beyond the chiral
limit, this structure is clearly PS-PV coupling-scheme dependent.
Therefore, the successful reproduction of the empirical value for $g_{\pi N}$
may provide an important QCD constraint for 
the pion-nucleon interaction type.

To see the coupling scheme dependence more clearly,  
we use the PS and PV Lagrangians
\begin{eqnarray}
{\cal L}_{ps} = g_{\pi N} {\bar \psi} i \gamma_5 {\mbox {\boldmath $\tau$}}
\cdot {\mbox {\boldmath $\pi$}} \psi\;; \quad 
{\cal L}_{pv} = {g_{\pi N} \over 2m}  {\bar \psi} \gamma_5 \gamma_\mu
 {\mbox {\boldmath $\tau$}} \cdot \partial^\mu {\mbox {\boldmath $\pi$}}
\psi\ ,
\label{pspv}
\end{eqnarray}
in constructing the phenomenological side of the correlator, Eq.~(\ref{two}).
Then the correlator is expanded in terms of the pion momentum $p_\mu$.
Using the PS Lagrangian, we obtain for the $i\gamma_5$ structure~\cite{hung2},
\begin{eqnarray}
g_{\pi N} \lambda^2 \left [- {1 \over q^2-m^2} -
{p\cdot q \over (q^2 - m^2)^2} + {p^2 \over (q^2-m^2)^2}\right ]
+\cdot\cdot\cdot\ .
\end{eqnarray}
Here $\lambda$ is coupling of $J_N$ to the physical nucleon, $m$ is 
nucleon mass. Note that the first term is the phenomenological part of the
sum rule in the soft-pion limit~\cite{hat}.  The second term
containing $p\cdot q$ is not the same chiral order as $m_\pi^2$.
Thus at $p^2=m^2_\pi$, the phenomenological correlator takes the form,
\begin{eqnarray}
m_\pi^2 { g_{\pi N} \lambda^2 
\over (q^2 -m^2)^2}+ \cdot\cdot\cdot \ .
\label{ps}
\end{eqnarray}
The ellipses indicate the contribution when $J_N$ couples to higher resonances.
This includes the continuum contribution
whose spectral density is usually parameterized by a step function
with a certain threshold $S_\pi$, and single pole terms associated with
the transitions, $N\rightarrow N^*$~\cite{ioffe}.

On the other hand, with the PV Lagrangian, the similar recipe yields the
correlator at the order $m_\pi^2$,
\begin{eqnarray}
{m_\pi^2 \over 2} { g_{\pi N} \lambda^2 \over (q^2 -m^2)^2} +
\cdot\cdot\cdot\ ,
\label{pv}
\end{eqnarray}
Note that in the PV case, there is no soft-pion limit as it should be.
This PV correlator contains an additional 
residue of $1/2$ compared to the
PS correlator. Thus, $g_{\pi N}$ determined from the PV coupling
scheme is twice of the one from the PS coupling scheme.

In the construction of this sum rule, the pion mass, $m_\pi^2$, 
will be taken out as an overall factor.  The rest correlator will
be used to construct the sum rule.  Then, a consistent treatment
should be made also in the OPE side. Namely, from   
the Gell-Mann$-$Oakes$-$Renner relation,
\begin{equation}
-2 m_q \langle {\bar q} q \rangle = m_\pi^2 f_\pi^2\ ,
\end{equation}
the vanishing limit of the pion
mass, $m_\pi^2 \rightarrow 0$, is consistent with the chiral limit,
$m_q \rightarrow 0$.
Therefore, for the sum rule with $m_\pi^2$ taken out as an
overall factor, the quark-mass term should be kept in
the OPE side.   Clearly, this aspect has been overlooked in our 
previous calculations~\cite{hung2} and needs to be implemented.

To construct the OPE side, we consider the correlation function with a
charged pion,
\begin{eqnarray}
\Pi (q,p) = i \int d^4 x e^{i q \cdot x} \langle 0 | T[J_p (x)
{\bar J}_n (0)]| \pi^+ (p) \rangle\ .
\label{two2}
\end{eqnarray}
Here $J_p$ is the proton interpolating field suggested by Ioffe~\cite{ioffe},
\begin{eqnarray}
J_p = \epsilon_{abc} [ u_a^T C \gamma_\mu u_b ] \gamma_5 \gamma^\mu d_c\ ,
\end{eqnarray}
and the neutron interpolating field $J_n$ is obtained by replacing
$(u,d) \rightarrow (d,u)$.
In the OPE,  we keep the quark-antiquark component of the
pion wave function
and use the vacuum saturation hypothesis
to factor out
higher dimensional operators in terms of the pion wave function and the
vacuum expectation value.

For the sum rule with the $i\gamma_5$ structure, we replace the 
quark-antiquark component 
of the pion wave function as follows,
\begin{eqnarray}
\langle 0 | u^\alpha_a (x) {\bar d}^\beta_{a'} (0) | \pi^+ (p) \rangle\ 
\rightarrow
{\delta_{a a'} \over 12 } (i \gamma_5)^{\alpha \beta}
\langle 0 |
{\bar d}(0) i \gamma_5  u (x) | \pi^+ (p) \rangle\ .
\label{diquark}
\end{eqnarray}
At $p^2=m_\pi^2$ order, the matrix element in the left-hand side
is replaced as~\cite{bely}
\begin{eqnarray}
\langle 0 |
{\bar d}(0) i \gamma_5  u (x) | \pi^+ (p) \rangle\ 
\rightarrow -m^2_\pi {\sqrt{2} \langle {\bar q} q \rangle \over 3 f_\pi}\ ,
\end{eqnarray}
where the overall normalization of the pion wave function at the second
moment has been used.
Another contribution at $m_\pi^2$ order is  obtained by
moving a gluon tensor from a quark propagator into the quark-antiquark
component.  This constitutes the three particle
wave function whose overall normalization is relatively well-known.
From Ref.~\cite{bely}, 
\begin{eqnarray}
\langle 0 |G_{\mu\nu}^A (0) u^\alpha_a (x) 
{\bar d}^\beta_b (0) |\pi^+ (p)\rangle
=- {i f_{3\pi} \over 32 } m_\pi^2 t^A_{ab} 
(\gamma_5 \sigma_{\mu\nu})^{\alpha \beta}\ ,
\label{gluon}
\end{eqnarray}
where~\footnote{Its value
is uncertain by an error $\pm 0.0005$ GeV$^2$ depending on the
renomalization scale~\cite{bely1}.  However, the contribution from
Eq.~(\ref{gluon}) is small in our sum rule as we will discuss below.
Thus, this error in $f_{3\pi}$ is negligible in our sum rule.}
$f_{3\pi}=0.003$ GeV$^2$
and the color matrices $t^A$ are related
to the Gell-Mann matrices via $t^A=\lambda^A/2$.

As we have discussed, the linear terms in quark mass should be
kept in the OPE for the sum rule at $m_\pi^2$ order.  
The quark-mass dependent terms can be obtained by 
first taking the limit, $p_\mu \rightarrow 0$, in the quark-antiquark 
component, Eq.~(\ref{diquark}),
while picking up linear terms in quark-mass from the other part of the 
correlator~\footnote{For complete quark propagator including the 
linear order in 
quark mass, see Ref.~\cite{wilson}. Note that gluonic tensor used there
has opposite sign of that in Ref.~\cite{qsr}. This is just a matter of how one
defines the covariant derivative but, in practice, 
this sign difference should be carefully noted.}.
It turns out that the condensates, $m_q\langle {\bar q} q \rangle$ and
$ m_q \langle {\bar q} g_s \sigma \cdot {\cal G} q \rangle \equiv
m_q m_0^2 \langle {\bar q} q\rangle$, 
contribute to the OPE of the $i\gamma_5$ structure.
The Gell-Mann$-$Oakes$-$Renner relation is used to convert
$m_q\langle {\bar q} q \rangle$ to 
$-m^2_\pi f_\pi^2/2$.   
Therefore, the quark-mass terms give additional contributions
to the sum rule at $m_\pi^2$ order.

Collecting all the OPE terms contributing to the $i\gamma_5$
structure at $m_\pi^2$
order,  the OPE side (after taking out the isospin factor $\sqrt{2}$ as
well as $m_\pi^2$ as overall factors)
takes the form
\begin{eqnarray}
ln(-q^2) \left [ {\langle {\bar q}q \rangle \over 12 \pi^2 f_\pi}
              + {3 f_{3\pi} \over 4\sqrt{2}\pi^2} \right ]
+  f_\pi \langle {\bar q}q \rangle {1\over q^2}
+ {1\over 72 f_\pi} \langle {\bar q}q \rangle 
\left \langle {\alpha_s \over \pi} {\cal G}^2
\right \rangle {1\over q^4}
-{1\over 3} m_0^2 f_\pi \langle {\bar q}q \rangle {1\over q^4}
\end{eqnarray}
Note, we use the pion decay constant $f_\pi =0.093$ GeV here.  
The second and fourth terms come from the quark-mass dependent terms.
It turns out that these are important in stabilizing the sum rule,
justifying the inclusion of quark-mass terms in the OPE.
The second term in the bracket comes from gluonic contribution combined
with the quark-antiquark component, Eq.~(\ref{gluon}).  Its contribution
is about 4 times smaller than the first term in the bracket.
Except for this term, all others contain the quark condensate.
This feature provides very stable results when this sum rule
combined with the nucleon chiral-odd sum rule.

We now match the OPE with its pseudoscalar 
phenomenological part, Eq.(\ref{ps}).  
To saturate the correlator with the low-lying resonance, 
we perform Borel transformation and obtain,
\begin{eqnarray}
&&g_{\pi N} \lambda^2 e^{-m^2/M^2} [ 1+ AM^2]= 
\nonumber \\
&&
-M^4 E_0 (x_\pi) \left [ 
{\langle {\bar q}q \rangle \over 12 \pi^2 f_\pi}
              + {3 f_{3\pi} \over 4\sqrt{2}\pi^2} \right ]
- f_\pi \langle {\bar q}q \rangle M^2 
+ {1\over 72 f_\pi} \langle {\bar q}q \rangle
\left \langle {\alpha_s \over \pi} {\cal G}^2\right \rangle
-{1\over 3} m_0^2 f_\pi \langle {\bar q}q \rangle\ .
\label{sum1}
\end{eqnarray}
The contribution from $N \rightarrow N^*$~\cite{ioffe} is denoted by 
the unknown constant, $A$. The continuum contribution is included in 
the factor, $E_n(x_\pi \equiv S_\pi /M^2)= 1- (1 + x_\pi +
\cdot\cdot \cdot + x^n_\pi/n!) e^{-x_\pi}$
where $S_\pi$ is the continuum threshold, which we take 2.07 GeV$^2$
corresponding to the Roper resonance. In our analysis, we
take standard values for the QCD parameters,
\begin{eqnarray}
\langle {\bar q} q \rangle &=& -(0.23~{\rm GeV})^3
\;; \quad
\left \langle {\alpha_s \over \pi} {\cal G}^2 \right \rangle =
(0.33~{\rm GeV})^4\;; \quad m_0^2 = 0.8~{\rm GeV}^2\ .
\end{eqnarray}

In figure~\ref{fig1}, we plot $g_{\pi N} \lambda^2 [1 + A M^2]$
as a function of the Borel mass $M^2$. To see the sensitivity to the
continuum threshold, we also plot the curve with
$S_\pi =2.57$ GeV$^2$, which is very close to the one with 
$S_\pi =2.07$ GeV$^2$. The two curves differs only by 2\% at $M^2=1$ GeV$^2$,
indicating that our sum rule is insensitive to the continuum threshold.
The highest dimensional term in the OPE contributes appreciably for
$M^2\le 0.6$ GeV$^2$, more than 20 \% of the total OPE. 
Thus, the truncated OPE therefore will be reliable for $M^2 \ge 0.6$ GeV$^2$.
The slope of the curve for $M^2 \ge 0.6$ GeV$^2$ is small, 
indicating that the unknown single pole term denoted by $A$ is 
small. 

To eliminate the dependence on the unknown strength $\lambda$ in our
sum rule, we divide Eq.~(\ref{sum1}) by
the nucleon chiral-odd sum rule and obtain,
%
\begin{eqnarray}
&&{g_{\pi N}\over m}[ 1+ AM^2]=\nonumber \\
&&
\left \{ M^4 E_0 (x_\pi) \left [{1\over 3f_\pi} + {3 f_{3\pi}\over 
\langle {\bar q}q \rangle \sqrt{2} } \right ] + {8\pi^2 f_\pi\over 3} M^2
-{\pi^2 \over 18 f_\pi} \left \langle {\alpha_s \over \pi} {\cal G}^2
\right \rangle  + {4\pi^2\over 3} m_0^2 f_\pi \right \}
\nonumber \\
&&\times
\left \{ M^4 E_1 (x_N) - {\pi^2\over 6} 
\left \langle {\alpha_s \over \pi} {\cal G}^2
\right \rangle  \right \}^{-1}\ ,
\end{eqnarray}
where $x_N=S_N/M^2$ with $S_N$ being the continuum threshold for
the nucleon sum rule.  
Note that the dependence on the quark condensate has been mostly 
canceled in the
ratio, leaving a slight dependence in the term $f_{3\pi}$.
Additional source of the uncertainty associated with the
gluon condensate is also very small as it is canceled in the ratio.  
One important uncertainty comes from the parameter $m_0^2$, which
however appears only in the highest dimensional OPE. Therefore,
its contribution will be suppressed in the Borel window chosen.  
The error from QCD parameters is estimated
numerically  and it is about $\pm 1.2$ in determining $g_{\pi N}$. 
For the continuum threshold in the nucleon sum rule, we take $S_N = S_\pi$. 
This choice is made because at the chiral limit the $i\gamma_5$ sum rule
is equivalent to the nucleon chiral-odd sum rule; these two  
are related
by chiral rotation~\cite{krippa1}. This equivalence provides
the Goldberger-Treiman relation with $g_A=1$~\cite{hat}. 
This choice for the continuum is also supported 
from modeling higher resonance contributions to the correlator
based on effective models~\cite{hung2}.  
We determine $g_{\pi N}$ and
$A$ by fitting the RHS with a straight line within the appropriately
chosen Borel window.  The dependence on the Borel mass is mainly driven
by the nucleon sum rule. 
The maximum Borel mass is determined by restricting the the continuum 
contribution 
from the nucleon sum rule while the minimum Borel mass
is obtained by restricting the highest OPE term 
from the $\pi NN$ sum rule. These gives the common window of the
two sum rules, $0.65 \le M^2 \le 1.24$.
By fitting the RHS with a straight line within this window, we obtain
$g_{\pi N} = 13.3 \pm 1.2$, where the quoted error comes from 
the QCD parameters.  This is remarkably close
to its empirical value of 13.4.

In getting this result, it is essential to go beyond the chiral limit.
Since the empirical $g_{\pi N}$ should include the chiral corrections,
it is indeed natural to go beyond the chiral limit in the determination of 
$g_{\pi N}$. One important observation made in this work
is that the quark-mass dependent
terms shouldn't be treated separately from the sum rule proportional
to $m_\pi^2$ as they constitute the same chiral order via
the Gell-Mann$-$Oakes$-$Renner relation.  The quark-mass terms are
found to be
very important in stabilizing the sum rule. 
Our findings, remarkable agreement with the
empirical value and the insensitiveness on the QCD parameters,
may provide a solid ground in constructing 
sum rules for other meson-baryon couplings.  The predictive
power of QCD sum rules can be substantially enhanced.  One 
application of our sum rule to the $\eta NN$ 
coupling is in progress~\cite{hung3}.  
Furthermore, our result provides a QCD 
constraint for the type of the pion-nucleon coupling. The value
of $g_{\pi N}$ quoted above is based on the PS coupling scheme.
With the PV coupling scheme, we would have obtained the value twice
of the quoted above. [See Eq.(\ref{pv}).]  Any error in our
approach  can not produce  
the value of $g_{\pi N}$ consistent with the PV coupling scheme.  
Therefore our work 
suggests that the PS scheme is preferable for the pion-nucleon
coupling from the QCD point of view.

In summary, we have developed a QCD sum rule for $\pi NN$
coupling beyond the chiral limit
for the first time.  The quark-mass dependent terms are
combined to the sum rule proportional to $m^2_\pi$ and 
they are very important in this sum rule. 
A remarkable agreement with the empirical value of $g_{\pi N}$ 
was obtained with very small errors.  
This sum rule provides the first QCD constraint for the
type of the pion-nucleon coupling, in favor of the pseudoscalar coupling.

\acknowledgments
The author is indebted to T. Hatsuda who has drawn the attention to
this problem.  The author also thanks M. Oka, and S. H. Lee
for useful discussions.
This work is supported by Research Fellowships of
the Japan Society for the Promotion of Science.

\begin{figure}
\caption{ The Borel mass dependence of $g_{\pi N} \lambda^2 [1 + A M^2]$.
The solid line is for $S_\pi = 2.07 $ GeV$^2$ and the
dashed line is for $S_\pi =2.57 $ GeV$^2$. The two
curves are differed only by 2\% at $M^2=1$ GeV$^2$.}
\label{fig1}
\end{figure}

\setlength{\textwidth}{6.1in}   
\setlength{\textheight}{9.in}  
\begin{figure}
\centerline{%
\vbox to 2.4in{\vss
   \hbox to 3.3in{\includegraphics{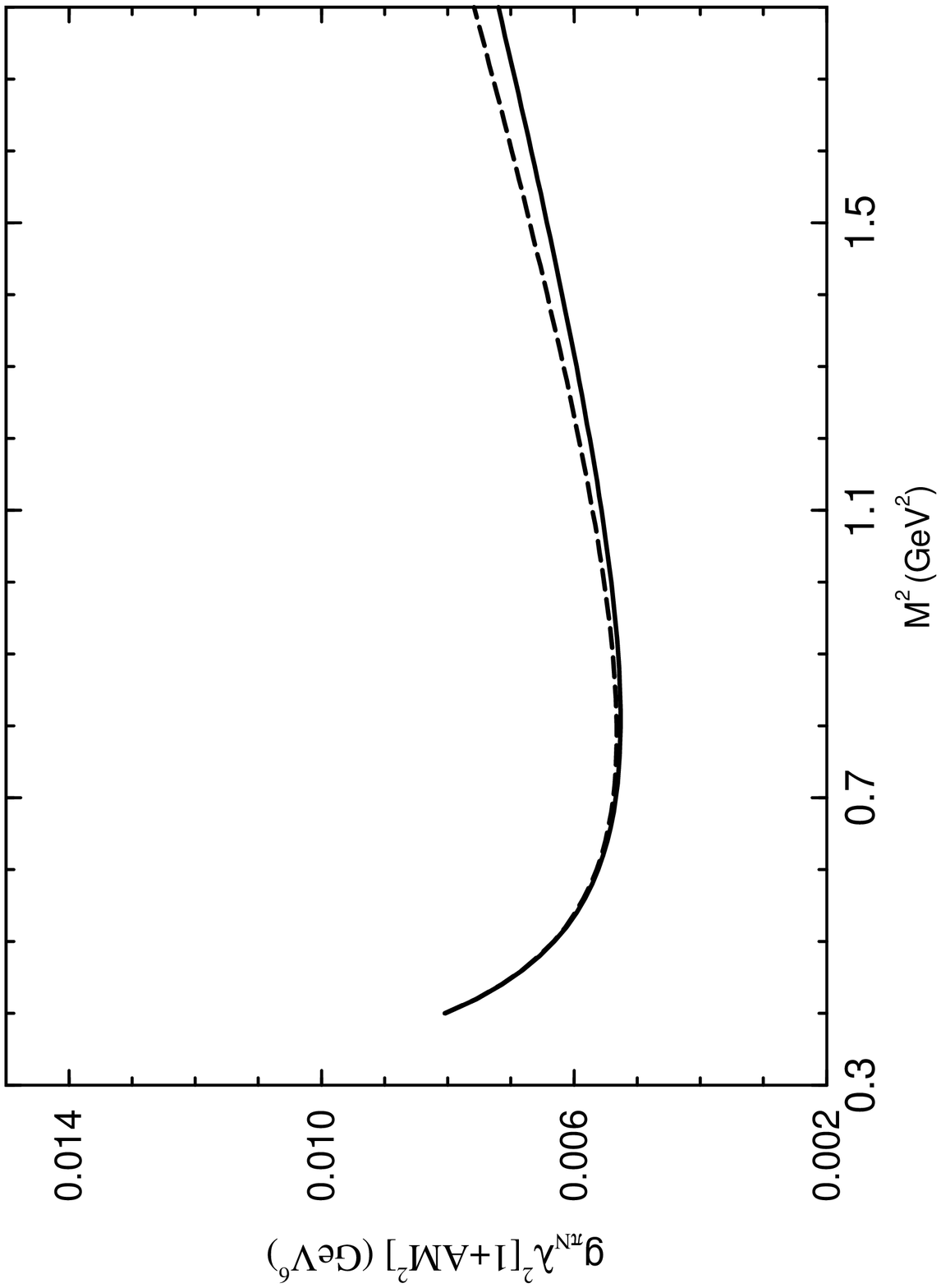}\hss}}
}
\bigskip
\vspace{400pt}
\end{figure}


\begin{references}
\bibitem{SVZ}     {M.A. Shifman, A.I. Vainshtein, and V.I. Zakharov,
                            Nucl. Phys. {\bf B 147} (1979) 385,448.}
\bibitem{qsr}     {L.J. Reinders, H. Rubinstein and
                    S. Yazaki, Phys. Rep. {\bf 127} (1985) 1.}

\bibitem{hat}     {H. Shiomi and T. Hatsuda,
                            Nucl. Phys. {\bf A 594} (1995) 294.}
\bibitem{krippa1}     {M. C. Birse and B. Krippa,
                            Phys. Lett. B {\bf 373} (1996) 9.}
\bibitem{hung1}     {Hungchong Kim, Su Houng Lee and Makoto Oka,
                           Phys. Lett. B {\bf 453} (1999) 199.}
\bibitem{hung2}     {Hungchong Kim, Su Houng Lee and Makoto Oka,
                        Los Alamos Preprint, nucl-th/9811096,
                        {\it To be published in Physical Review D.};
                     Hungchong Kim, Su Houng Lee and Makoto Oka,
                        Los Alamos Preprint, nucl-th/9902031.}
\bibitem{bely1}   {V. M. Belyaev and Ya. I. Kogan,
                   JETP Lett. {\bf 37}, (1983) 730;
                   B. L. Ioffe and A. G. Oganesian,
                   Phys. Rev. D {\bf 57} (1998) R6590.}
\bibitem{ioffe}     {B. L. Ioffe and A. V. Smilga,
                            Nucl. Phys. {\bf B 232} (1984) 109.;
                 B. L. Ioffe,
                            Nucl. Phys.  {\bf B188} (1981) 317.}
 
\bibitem{bely}     {V. M. Belyaev, V. M. Braun, A. Khodjamirian and R. R\"uckl,
                            Phys. Rev. D {\bf 51} (1995) 6177.}
\bibitem{wilson}     {J. Pasupathy, J. P. Singh, S. L. Wilson and C. B. Chiu,
                            Phys. Rev. D {\bf 36} (1987) 1442.;
                      S. L. Wilson, Ph.D thesis, 
                      University of Texas at Austin, 1987.}
\bibitem{hung3}  { Hungchong Kim, {\it In preparation}}

\end{references}
\end{document}